\documentclass[journal,comsoc]{IEEEtran}
\usepackage{amsmath,amssymb,amsfonts,amsthm}
\usepackage{algorithmic}
\usepackage{graphicx}
\usepackage[flushleft]{threeparttable}
\usepackage{color}
\usepackage{cite}

\begin{document}
\title{Preventing the attempts of abusing cheap-hosting Web-servers for monetization attacks}
\author{\IEEEauthorblockN{Van-Linh Nguyen, Po-Ching Lin, and Ren-Hung Hwang}\\
}

	
	\maketitle

	\begin{abstract}
		
		Over the past decades, the web is always one of the most popular targets of hackers. Today, along with the popular usage of open sources such as Wordpress and Joomla, the explosion of the vulnerabilities in such frameworks causes the websites using them to face numerous security threats. Unfortunately, many clients and small companies may not be aware of these serious security threats and call a rescuer only when the website is hacked, compromised, or blocked by the search engines. In this paper, we present an effective counter against such threats, including monetization attempts in the less valuable targets such as small websites.
		
	\end{abstract}
	
	\begin{IEEEkeywords}
		Web security attacks, Web vulnerabilities, Web protection.
	\end{IEEEkeywords}
	
	\IEEEpeerreviewmaketitle
	
	\section{Introduction}
	\label{chap:intro}
	
    According to Internet Live Stat \cite{internetlivestats}, the total number of online websites has already soared over 1.9 billion and is steadily increasing. The most popular content management systems (CMS) installed on such websites are PHP-based frameworks such as Wordpress and Joomla, which take, as of Dec. 2018, 59\% and 6\% of the market share \cite{w3tech,trends}. Note that a large number of those websites belongs to small companies and individuals that may have few motivations to pay thousands of dollars for security maintenance services. Also, due to the development cost, the owners, even the developers they hire, may prefer available CMS such as Joomla to accelerate the deployment. However, that presents serious security threats, in which the website can be the target of large scale exploitation campaigns. For example, an attacker can exploit many websites with the same vulnerabilities of the same framework. The hardware-based firewalls such as \cite{Citrix} and scrubbing centers such as CloudFlare may have few chances to reach the clients who are only willing to pay several dollars per month for basic web services in shared hosting. On the other hand, the open source firewall such as Mod Security \cite{modsecurity} can be available in shared hosting but have few mechanisms to prevent all the attacks against the vulnerabilities of CMS core and its add-ons. As usual, the lack of security protection and maintenance in these websites plays an important role to motivate the cybercriminals to attack and launch illegitimate activities. In this case, those are the monetization attempts. 
    
    
  
   \textbf{The era of all for the money}
   
     A long time ago, attacking a web can be a demonstration of hacking skills, e.g., for script kiddies. Then there are increasingly sophisticated attacks to steal the client data of online shopping websites, although such important systems are often reinforced by good solutions and the owner may not be interested in using open sources. Nowadays, while the small websites such as personal blogs may have little valuable information to steal, we have discovered a campaign, in which attackers have abused the hacked websites to make money. Mining cryptocurrency is an example. Coinhive allows website owners to make use of their website visitors' CPU to mine the Monero cryptocurrency \cite{monero}. Following that, the attacker targets to inject mining scripts (cryptojacking) to as many victims as possible and then launch long-term silent monetization campaigns \cite{Tommy12}. 
     
     That is not the end of the monetization chain of the attacker. The malicious scripts in an injected website can be used to force the visitors to redirect to a fake address, force them to install given shareware software, open pop-up ads and so on. Attracted by high CPC (cost per click) rates \cite{cpckeyword}, the attacker can hijack the websites, replace the content of a page and entice the visitors to click on the fake links \cite{Tommy12}. We have also found such silent attempts in building a farm web for phishing and ads campaign through an effort of rescuing injected websites. 
     
      In summary, this article makes the following contributions: 
     
     \begin{itemize} 
     	
     	\item We discuss the detail of the vulnerabilities exploitation, monetization attempts, and Web-server abuse. This attempt is driven by the security characteristics of the websites based on the cheap hosting but accounts for the majority of web users. 
     	
     	\item We unveil a deep insight into our cleaning and attack prevention mechanism to crack down the attempts of abusing cheap-hosting Web-servers for monetization attacks.
     	
     	\item We practically implement the proposed system and rescue several websites and the servers which were hacked and abused.
    
    \end{itemize}
     
     The remainder of this paper is organized as follows. Section~\ref{chap:backgroundattack} covers the detail of those attack attempts. Section~\ref{chap:proposal} then discuss a systematic approach to clean and prevent such attacks. Section~\ref{chap:evaluation} covers the implementation in practice. To conclude, we discuss the goal of the review in Section~\ref{chap:conclusion}.

	\section{Attack tactics and examples}
	\label{chap:backgroundattack}
	
	\subsection{Attack tactics}
	\label{attacktactic}
	
\begin{figure}[ht]
		\begin{center}
			\includegraphics[width=1\linewidth]{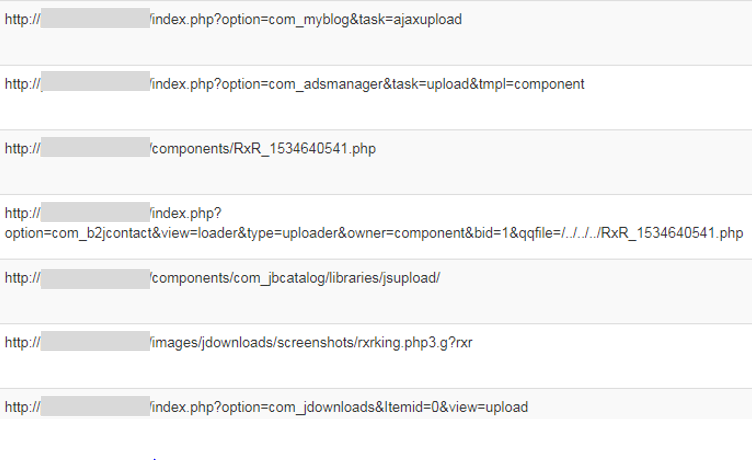}
		\end{center}
		\caption{Illustration of an attack with many exploitation calls, including targeting potential add-ons installed on the CMS, if any. }
		\label{fig:exploitation-code}
	\end{figure}

	 Exploiting the vulnerabilities to compromise the website is often the first step but important for further sophisticated campaigns such as phishing and monetization. There are several options to attack a web; however, the most common approach is taking advantage of exploiting zero-day and well-known vulnerabilities. Through the system log in the rescue website, we have found that the attacker used an automated exploitation script to try various attacks (as illustrated in Fig.~\ref{fig:exploitation-code}) was successful with a famous vulnerability of SQLi in Joomla Core 3.7.x in a website and a remote execution code vulnerability in updating the style of the default template protostar in another site. 
	 
	 
	 In practice, designing an auto-attack script with many exploitation patterns \cite{codexploit} is one of the most effective methods to increase the probability of a successful attack. Besides, the reality of many websites using the old versions of CMS core is not uncommon, and this fragmentation makes the developers harder to maintain security patches for many vulnerabilities at the same time. Another potential attack is to exploit the vulnerabilities of the add-ons or additional components installed to the CMS core, e.g., Wordpress Super Cache. The nulled commercial scripts also account for a significant part in worsening the security of a website. The newbie developers often use these shared scripts for saving development cost, but may not be aware of the malicious codes plugged inside. 
	 
\begin{figure}[ht]
		\begin{center}
			\includegraphics[width=1\linewidth]{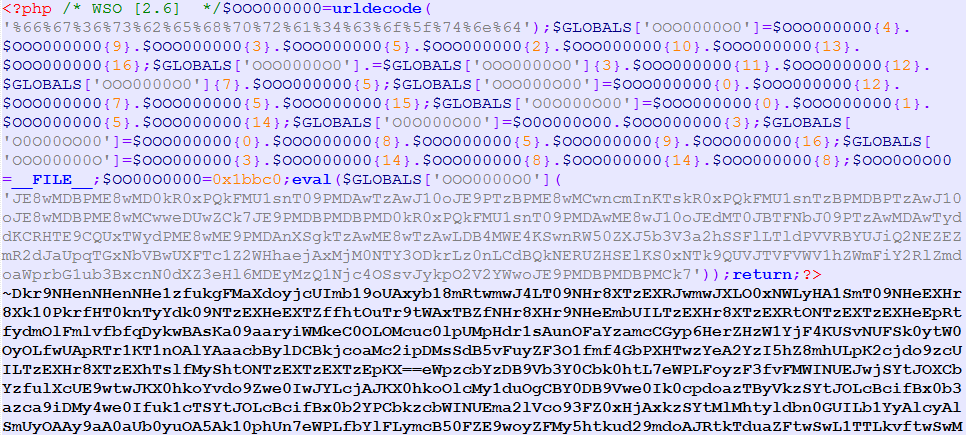}
		\end{center}
		\caption{Illustration of an obfuscated shell-code}
		\label{fig:shell-code}
	\end{figure}

	 The final target of the attacks mentioned above is mostly to upload a shell-code or attack script and then clone it to many folders of the web directory, normally with the possible unharmed names such as functions.php and libraries.php. A shell-code is an obfuscated script to be used to create a backdoor for the attacker's future access and do what he orders. Fig.\ref{fig:shell-code} illustrates the example of such shell-code. For the attack, in an effort of hiding the real IP address, the attacker uses many C\&C servers with different IP addresses from many countries to launch the attacks as well as deliver the updates to this shell.
	
   After having a shell-code or an attack script on the website, the attacker has many options in launching further abuses to the web servers \cite{Seyed13,Zarras17}. In our case, we reveal three examples of such exploitations as follows.
	
	\subsection{Cryptojacking}
	\label{cryptojacking}
\begin{figure}[ht]
		\begin{center}
			\includegraphics[width=1\linewidth]{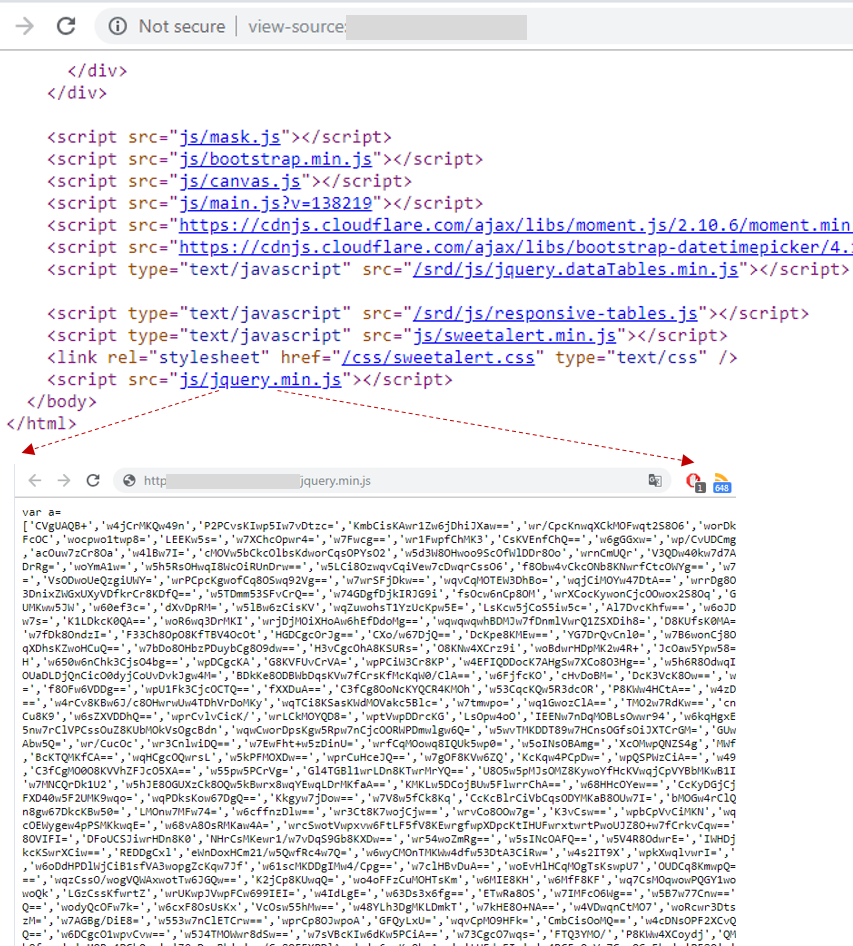}
		\end{center}
		\caption{An example of a crypto mining script}
		\label{fig:coins}
	\end{figure}

	Coinhive first introduced the crypto mining script function, a service that allows website owners to make use of their website visitors’ CPUs to mine the Monero cryptocurrency \cite{monero}. Coinhive introduced the crypto mining script as an alternative to placing ads on a website. The hacker often selects the cryptocurrency Monero for running crypto mining scripts due to its CPU-friendly hashing algorithm. Moreover, this cryptocurrency is anonymous and secure nature. In our case, the hacker injected several packed javascript scripts for the mining purpose with confusing names such as jquery.min.js, as shown in Fig.~\ref{fig:coins}.

While Monero is one of the first choices in this exploitation; other cryptocurrencies can be used as part of a crypto mining script too. The mining scripts can be interrupted if the visitors close their browsers; however, with small pop-ups (the user may not notice), the mining script is still running. To deal with this kind of attack, a possible approach is to find and clean all compromised codes or install the crypto mining script blockers. The detail of the first approach presents in the next section.

\subsection{Phishing attack}
	\label{phishing}
	
	The phishing attack is one of the most popular attacks and the target of many kinds of research and public anti-phishing scanning services such as Google Safe Browsing and Phishing tank \cite{phishtank}. Besides active scanning from these crawlers to detect the phishing websites, the anti-phishing services can rely on the manual reports of volunteers. However, that report is not always timely. In practice, the attacks often got caught when there is a significant increase in network traffic or the browser and search engines block the site. 
	
\begin{figure}[ht]
		\begin{center}
			\includegraphics[width=1\linewidth]{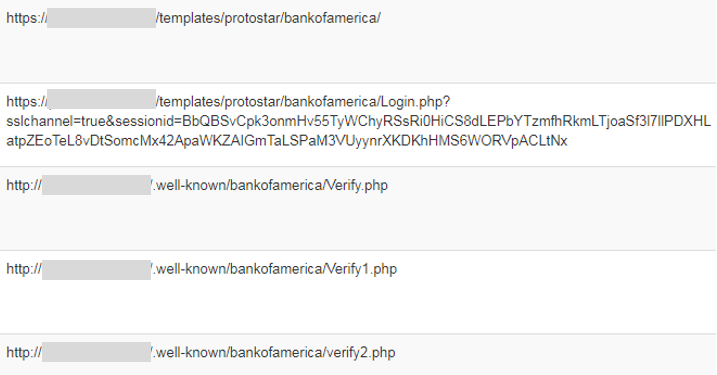}
		\end{center}
		\caption{An example of simple phishing tactic through faking the login form of the Bank of America}
		\label{fig:codephishing}
	\end{figure}

	With newly infected websites in hand, the attacker can carefully design a suitable phishing tactic to maximize efficiency. An example of the simple phishing tactic illustrated in Fig.~\ref{fig:codephishing}
	
   	\begin{figure}[ht]
   	\begin{center}
   		\includegraphics[width=1\linewidth]{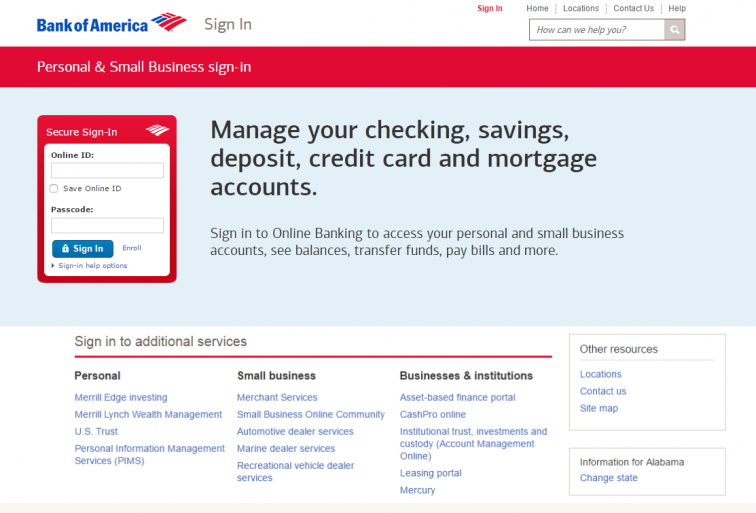}
   	\end{center}
   	\caption{Fake login page of the Bank of America website at the time of this writing. It may be trivial for the attacker to make a change to the page that follows the updates of the bank site. (the access link can be seen in Fig.~\ref{fig:codephishing})}
   	\label{fig:bankofamerica}
   \end{figure}

\begin{figure}[ht]
       	\begin{center}
       		\includegraphics[width=1\linewidth]{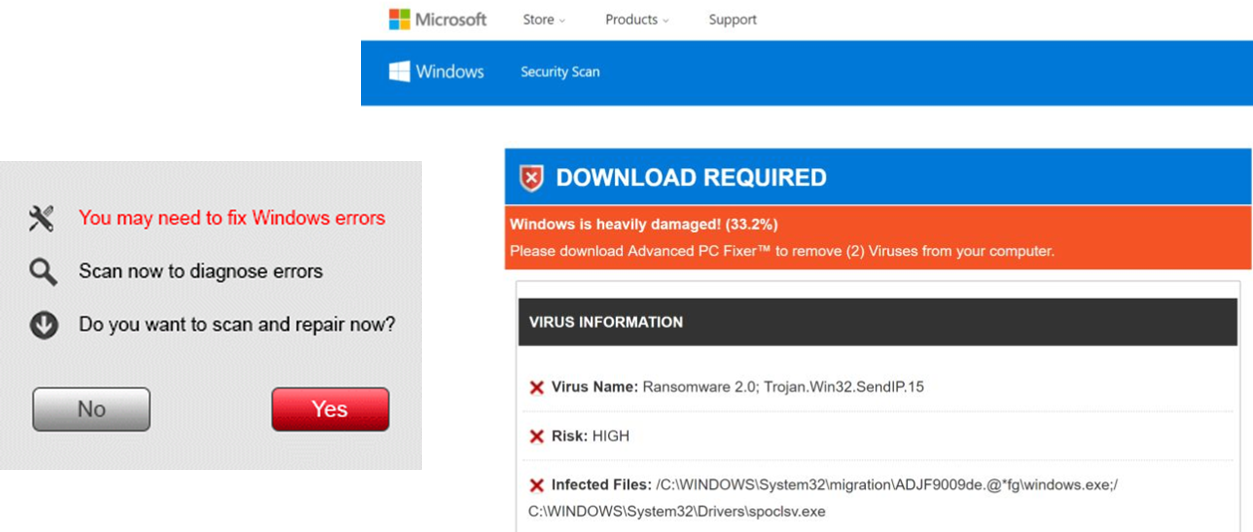}
       	\end{center}
       	\caption{(left)Illustration of a popup ads to warn the visitors for slowing PC and suggest to download a speedup PC software; (right) Illustration of an adware to guide for download PC Fixer, a fake program}
       	\label{fig:popup}
       \end{figure}

   On the other hand, the attacker can launch a PayPal Phishing using JavaScript redirect \cite{paypalphishing} in the online shopping-cart websites to capture credit card information. Even that, we found a malicious code tried to replace the domain name of the payment gateway in the API library. The other kind of popular phishing attacks found in our injected websites is to harass the visitors with warnings (adware), as illustrated in Fig.~\ref{fig:popup}, or auto download the files with a virus inside.
       
   \subsection{Fraud ads campaign}
   \label{fraudads}
   
   One of the most profitable industry is online advertising, in which Google accounts for the majority of click ads. The high profit of CPC keywords  \cite{cpckeyword} motivates the attacker to take advantage of loopholes to fraud the ads. While the advertisers always try to fix the loopholes and ignore the fraud ads, many fraud tricks may be not easy to detect, including the following one. First, due to having full remote access the web server (via the plugged shell-code), the attacker can analyze the available sitemap of the website and replace it with a manufactured version. Without frequent checking, the owner may not know this change. For the sites verified Google Sitemap, the attacker can periodically update the sitemap with their links and thus redirect the search content with their designed keywords.
   
\begin{figure}[ht]
   	\begin{center}
   		\includegraphics[width=1\linewidth]{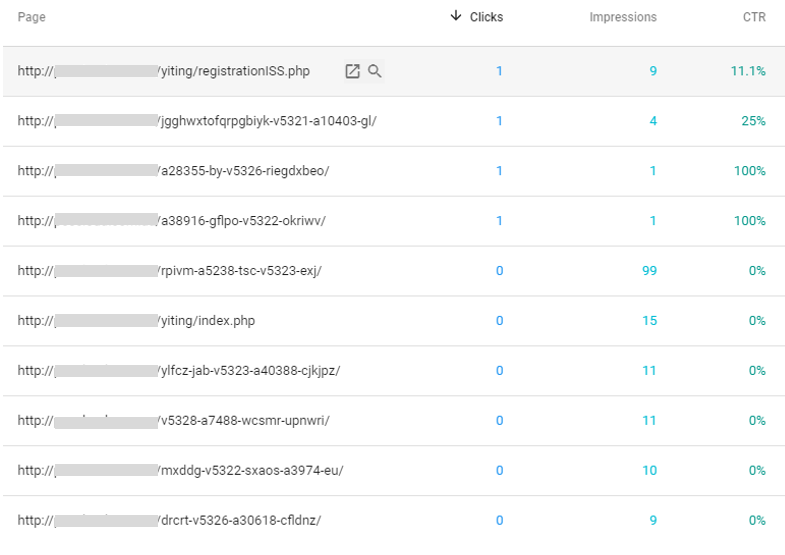}
   	\end{center}
   	\caption{An example of the sitemap updated by the attacker.}
   	\label{fig:ads1}
   \end{figure}

\begin{figure}[ht]
   	\begin{center}
   		\includegraphics[width=1\linewidth]{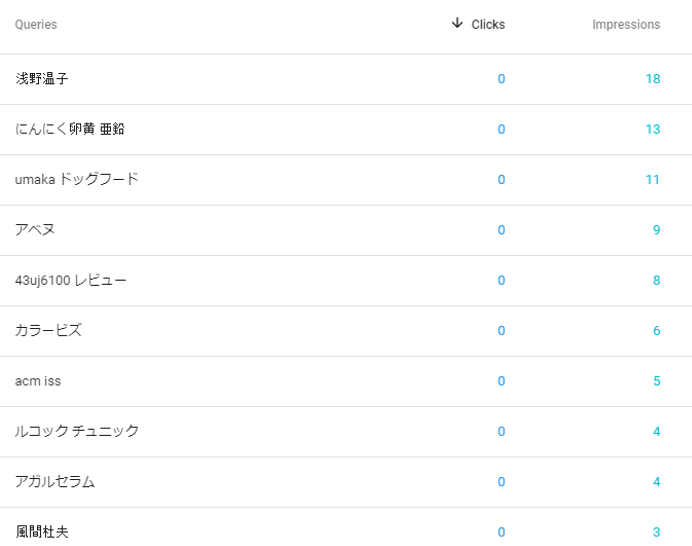}
   	\end{center}
   	\caption{The keywords and click results of the campaign (monitored in an hour) }
   	\label{fig:ads2}
   \end{figure}

   An example of such replacement can be seen in Fig.~\ref{fig:ads1}, where the attack bot generates several random links. The cybercriminals are often interested in creating links and content with profitable keywords such as pharma. The keywords and the click results of such a campaign can be seen via Google Console, as shown in Fig.~\ref{fig:ads1} and Fig.~\ref{fig:ads2}, respectively. According to our analysis, that campaign created tremendous traffic in a short time. That may not only potentially benefit the attacker’s pocket but also attract attention. In this case, that is us. 
   
\begin{figure}[ht]
	\begin{center}
		\includegraphics[width=1\linewidth]{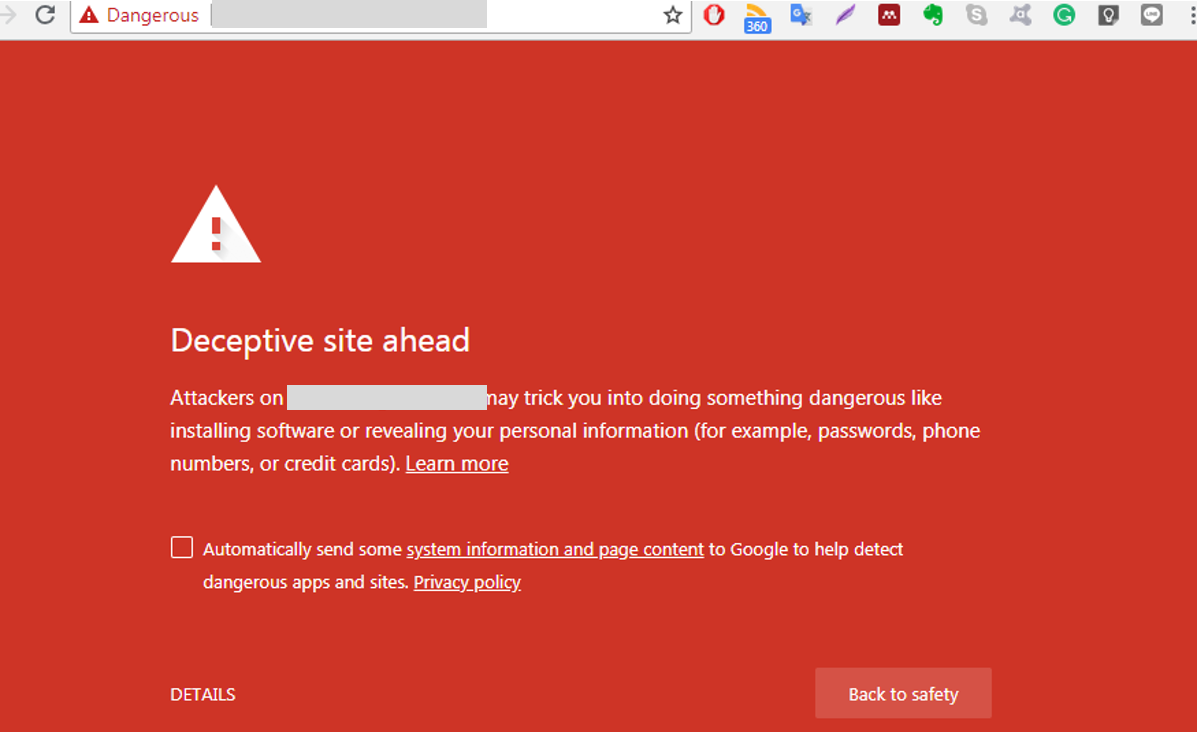}
	\end{center}
	\caption{Deceptive warning for visiting the website.}
	\label{fig:ads3}
\end{figure}

   Finally, the most fear of the owners is that the search engines block their websites, as illustrated in Fig.\ref{fig:ads3}. To restore these websites, the owners need to clean up (or hire a developer to do) first and then request a review to the anti-phishing services, e.g., via Google Webmaster. It may take days to verify the whole process, and folks may not want this situation to recur.

\section{Clean and Attack Prevention mechanism}
\label{chap:proposal}
\begin{figure}[ht]
  	\begin{center}
  		\includegraphics[width=1\linewidth]{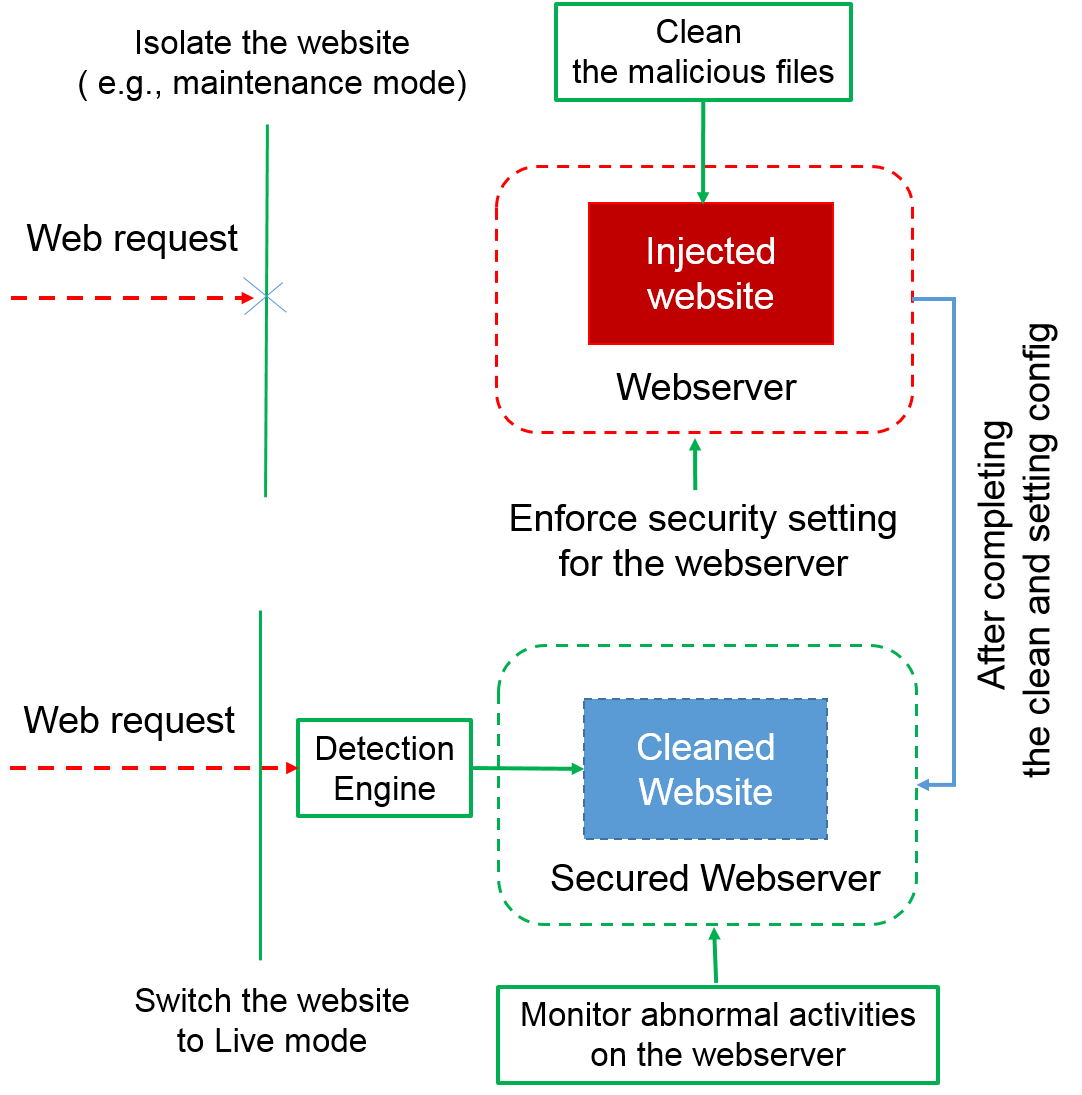}
  	\end{center}
  	\caption{The main functions of the security protection system}
  	\label{fig:architecture}
  \end{figure}

	For the small websites, it is infeasible to require high-end security protection such as \cite{Citrix}, primarily due to the economic reason. If the malicious scripts inject a site, the fastest way to restore the system function of the site is to use a backup version. The hosting providers can support a daily backup, but the owner must often pay an extra fee. In the case of having a backup of the latest version (due to no many changes to the site for a long time), it is possible to restore the original state when the website is stable. However, we argue that the backup tactic is only a temporary solution and does not solve the root problem because we cannot entirely guarantee whether the original version is malware-free. In this case, cleaning the injected files, i.e., removing the malicious codes, maybe the best and the most economical. Trusting the tactic, we deploy two steps to clean and enforce affordable security protection on the website. Note that our system can directly run on shared hosting without requiring additional components such as specific hardware and a dedicated server. The primary functions of the system are illustrated in Fig.\ref{fig:architecture}.

 	The first module consists of three separate functions: (1) isolate the website (put it to the maintenance mode or require a special link with given code to login to the website) to analyze and clean malicious codes; (2) check the security configuration of the web server such as unsafe PHP function (e.g., $shell$\_$exec$) and directory permission (3) activate the security checking (detection) and the prevention engine. We only switch the website back to the production mode after completing all (1), (2), and (3). Intuitively, the three tasks are likely to separate in function but designed to run together simultaneously. The first task is essential to remove all malicious codes and stop broadcasting the phishing or mining scripts. The second is used to properly configure and reinforce the server security setting from potential harm, while the primary duty of the last one is to prevent the new cycles of the attacks.
 	
 	While the first module is designed to defend against the potential attacks from infiltrating the system, the second module is an independent script which periodically checks the behavior of the scripts on the server such as creating irregular links and files. The periodic check can be set by a cron job in shared hosting or a scheduler if the dedicated server is supported. Due to the intrinsic characteristic of the event-driven Web technology, verifying the request (the first module) is the passive approach to resist the compromised attacks. Meanwhile, the second module pro-actively keeps an eye on the system activity. This extensive monitoring is necessary because the granted Web requests (bypass the first module’s check) can be potentially malicious ones and scheduled to abuse the server.

\subsection{Clean up the injected website}
\label{chap:clean}

	To perform the cleanup task on an injected website, we first lock the whole unauthorized interaction from the Internet to the site. By turning into the maintenance mode, the risk of illegal communication such as spreading malicious codes to the visitors and storming email scams will be declined. Next, to detect malicious files, our system performs a file integrity check and signature-based scanning. The files in the fresh CMS core is first hashed and then a further check on the files of the injected website. Note that we cannot trust the code of the additional components, and so they are not hashed. In the check, if there is a file that does not match the hash code of the same file in the original core, it can be a malicious script. 
	
\begin{figure}[ht]
	\begin{center}
		\includegraphics[width=1\linewidth]{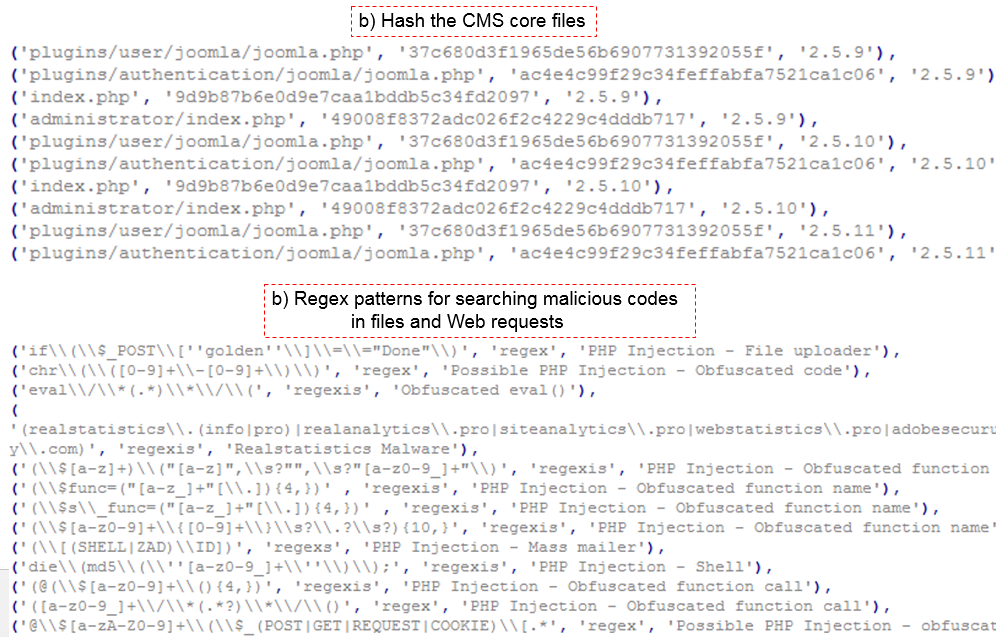}
	\end{center}
	\caption{Illustration of hashing code for the file integrity check and regex patterns for searching common malware.}
	\label{fig:signaturehash}
\end{figure}

	Meanwhile, another scanning is also launched on these files but using regex patterns, including variable obfuscation ( as shown in  Fig.~\ref{fig:signaturehash}. If found any pattern in those files, to be safe, the files will be blocked in execution and wait for a manual check by the administrator. Although scanning the integrity of CMS files may face a little challenge if the CMS core has been heavily modified to satisfy the new features that were not initially available on the website, we found that the scan with regex patterns (also namely signatures) is incredibly effective to support that missing part. The patterns we used in searching such common malware can be found in our repository \cite{repository}.    

\subsection{Reinforce the server security setting}
\label{chap:reinforce}

	The server security setting plays a critical role to significantly reduce the exploitation of a new cycle of the attacks. While several hosting providers may have a default policy to turn off the modules which cause potential security risks such as $shell\_exec$, $popen$, $proc\_open$ and $allow\_url\_include$, we can also do the same work through creating a custom $php.ini$ and $.htaccess$ in the root folder of the website. The setting of the website database such as the default “admin” or weak passwords are checked as well. Besides, the session lifetime and insecure permissions of file and directories are scanned and corrected with safe values. Note that activating HTTPS, SEO URLs, caching by default and periodically cleaning the temporary folder also definitely help to secure the website. 

\subsection{Attack Prevention}
\label{chap:prevention}

	The next important task is to detect new cycles of the attack attempts from the incoming Web requests and prevent them. Before conducting what is inside the Web requests, we first need to check whether the sender’s IP address is in any public blacklist. We do that through sending an IP validation request to the available servers of Tornevall DNS Blacklist, Spamhaus DNSBL, and Google Safe Browsing. Based on the return value, we consider the next conducting, e.g., blocking the access if that is the IP of a spam server. If the IP is valid, we perform extensive scanning on the received data (POST/GET) using the signatures similar to those of the clean-up process. Any suspicious access classified by this scan is temporarily blocked and waits for a further manual check by the administrator. In addition, a security warning page is also set to display. That will allow the requester to know what happens with their requests (e.g., in the case of false-positive detection). 
	
	To prevent the attempts of the attack script, the system denies all access from the User Agents generated by programming such as CURL. The high-risk extensions such as .php and .exe are also banned in the form submission, i.e., upload filtering. For multiple requests from the same IP, after three failed login attempts, the CAPTCHA is activated to prevent the automated hacks effectively. All the blocked attempts are traced for further analysis if necessary.
	
	For the website requiring a stricter protection policy, we can disable the features of installing new components/add-ons, or the owner must verify such installation requests manually. Another important policy is to disallow local and remote file inclusion. The former vulnerability, called directory traversal technique (such as controller=../../../etc/passwd), may allow an attacker to read sensitive files by exploiting poorly coded extensions while the latter one means the access to URLs (such as controller=http://www.virus.com/exploit.txt) that may let an attacker to download and run malicious scripts by exploiting poorly coded extensions.
	
	For the Web requests that target to mount a potential DDoS attack, the system activates the CAPTCHA after getting HTTP requests over a threshold in a short time, e.g., 200 per second. Only the access passing the CAPTCHA check can go ahead. Also, if the website is not a global site, i.e., targeting only domestic visitors, the blocking can prohibit all IPs that appear to be from selected countries. However, we must include exceptions of IPs for legitimate requests from servers such as search engines \cite{Algiryage18} and spam/blacklist check. The list of Geo IPs for the blocking-by-country function can be found at \cite{countryblocking}.
	
   \textbf{Blacklist explosion avoidance}
    Unlike distributed denial of service attacks (DDoS), the source used in this web attack may not exceed thousands, i.e., the IP which attacker used to evade the IP block. In practice, the attacker may launch the spoofing attacks, e.g., non-existing source IP, to overwhelm the blacklist storage. To defeat this exploitation, we can request a valid confirmation of the source and valid browser agent before accepting a post HTTP request. For the websites requiring high performance, we use a Bloom Filter \cite{Lin08} to query rapidly and memory-efficiently whether the hash of incoming HTTP request presents in the existing hash patterns. Other matching methods can be found at \cite{Sun18,Aditya17}.

\section{Eyeing on abnormal activities}
\label{chap:abnormal-activities}

	As we mentioned, the passive and data-centric searching system above relies much on detecting the malicious codes from the Web requests. However, several Web requests may bypass the scanning engine. In that case, an active monitor is essential to detect and isolate the spreading of malicious scripts timely. Thus, unlike the scanning process on the received data of the passive prevention engine, this script actively monitors the abnormal activity (behavior-centric) on the server. The events can be in a high frequency (over a threshold) of generating new links or external queries to the other IPs. Note that the monitor also keeps eyes on the actions of copying and modifying the core files, i.e., refer the hashed list.
	
	
\subsection{Identifying the Web-server abuse behavior and classification}
\label{chap:webserver-abuse}
     
 



 

\begin{figure}[ht]
  	    	\begin{center}
  	    		\includegraphics[width=1\linewidth]{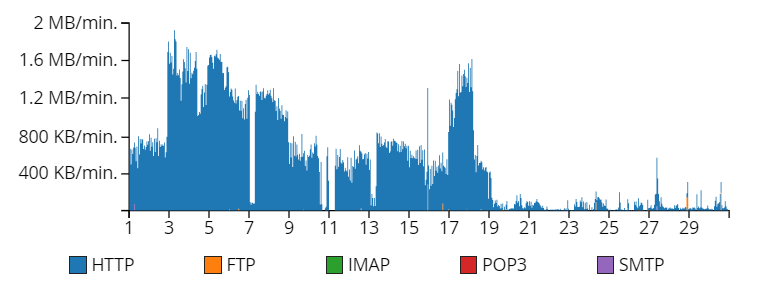}
  	    	\end{center}
  	    	\caption{The real-time monitoring of outgoing messages by protocols}
  	    	\label{fig:attack1}
    \end{figure}

  Each malicious activity often has several unique characteristics to detect. To differentiate between benign and malicious activities, we define the following features: the execute time of the script, the number of outgoing messages by protocols(e.g., SMTP). While the first feature can be used to monitor the resource abuse(long execution for a specific script), the second one is useful to catch the abnormal behavior of sending a large number of phishing emails (enormous SMTP requests). Fig.~\ref{fig:attack1} illustrates the real-time monitoring of such features. The hosting Cpanel may have a similar function in collecting these features. The additional patterns for analyzing HTTP requests can be found in \cite{Maria14}.
  
   After we extract the features, we need to classify whether they are malicious or benign. We use a decision tree, e.g., \cite{Hesham14}, with the features described above. The major advantage of using decision tree lies in the interpret-ability and human validation of the learned models. The advanced classification can be found in \cite{php-ml,google-ml}.

\subsection{Implementation}
	\label{chap:evaluation}

	Due to the limitation of intervention to the setting and server-script on the shared-hosting servers, particularly, Apache/Nginx, the system is preferably implemented in PHP but not limited to. Using the server-script language of the websites has significant advantages in deploying or integrating to the CMS (friendly-usage) or a standalone (if accessing the injected sites is impossible). For the system with more flexible control privilege, e.g., a dedicated server, the active and behavior-centric monitor is written in Python and can run as an independent process in memory.
	
\begin{figure}[ht]
		\begin{center}
			\includegraphics[width=1\linewidth]{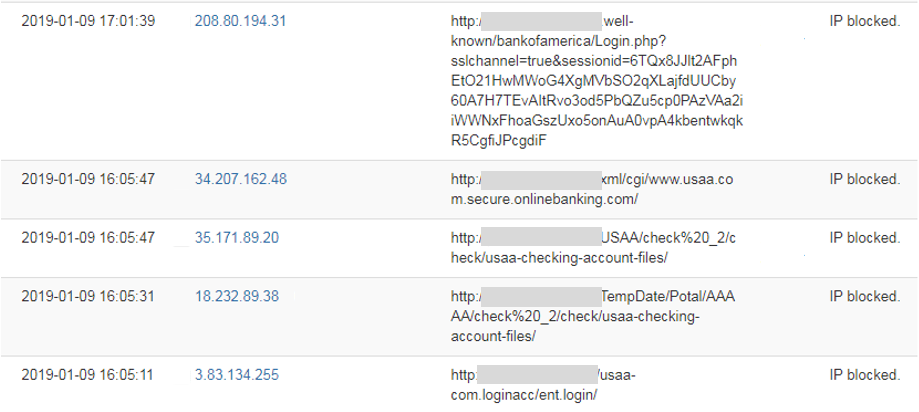}
		\end{center}
		\caption{The system can detect and prevent the attack attempts from various IP addresses}
		\label{fig:result1}
	\end{figure}

	As a result of the comprehensive approach, our system successfully rescues a set of injected websites and has effectively protected them. Fig.~\ref{fig:result1} shows that the automated attack script is still working and trying to attack our sites again but fails thanks to the reliable protection of our system. Because our website IPs are already stored in the attacker’s database at the first successful attack, new attack attempts will continue, unless the botnet handling these scripts is taken down.

	
\begin{figure}[ht]
   	\begin{center}
   		\includegraphics[width=1\linewidth]{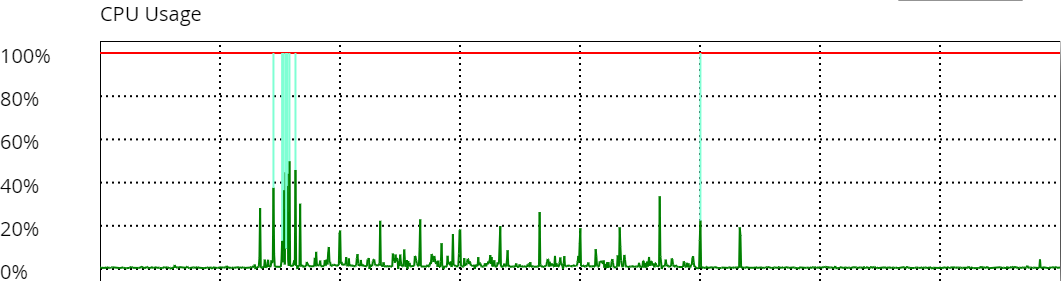}
   	\end{center}
   	\caption{The illustration of CPU Usage behavior monitoring. For the heavy executing scripts (high CPU usage), a detail notification of the path of the scripts and what they do will be sent to the administrator.}
   	\label{fig:cpuusage}
   \end{figure}

   Regarding the abnormal activity monitoring, the activities of the high Web server resource usage (as illustrated in Fig.~\ref{fig:cpuusage}) -- found by our classification -- are further checked whether they come from the scripts or the links inside the hashed list, including the original sitemap. If there is an exception, e.g., the heavy executing script is not in the hashed list or an abnormal increase of SMTP messages, a detail notification of the path of the file and what it does will be sent to the administrator.

	\section{Conclusion}
	\label{chap:conclusion}
	
    Through the above analysis, we are willing to demonstrate the attack attempts and tactics of the attackers in making money on compromised websites. A website that is blocked and not used for earning money does not mean the end of its life. There are probably thousands of such injected websites. We believe that as more such sites have been successfully taken down by the anti-phishing services and our security protection, the attacker is likely to change the attack tactics. The fight between the defender and the attacker can be considered as in a tit-for-tat battle, and the former player is not easy to win, i.e., preventing all the attacks. To not be lost in such struggle, our detection and prevention system is periodically updated with the new regex patterns and the hashed list for specific cases, e.g., the developer of the website may scan and confirm what files they have modified or what components to be added to the hashed list.  We hope that this revealing effort may benefit security favor readers, even place a brick of potential awareness of how the security rapidly shifts in world wide web.

\end{document}